\documentclass{PoS}
\usepackage{amsmath}
\newcommand{\PSfig}[2]{\includegraphics[width=#1]{#2}}
\newcommand{\vk}{{\bold{k}}}

\newcommand{\vkb}{{\bar{\bold{k}}}}
\newcommand{\vkt}{{\vk,\tau}}
\newcommand{\vkmt}{{-\vk,\tau}}
\newcommand{\vkbt}{{\bar{\vk},\tau}}
\newcommand{\vkmbt}{{-\bar{\vk},\tau}}

\title{
Auxiliary field Monte-Carlo study\\
of the QCD phase diagram at strong coupling
\thanks{Report No.: YITP-12-88, KUNS-2424}}

\ShortTitle{Auxiliary field MC study of QCD phase diagram at strong coupling}

\author{\speaker{Akira Ohnishi}\\
	Yukawa Institute for Theoretical Physics, Kyoto University,
	Kyoto 606-8502, Japan\\
        E-mail: \email{ohnishi@yukawa.kyoto-u.ac.jp}}

\author{Terukazu Ichihara\\
	Department of Physics,
	Kyoto University,
	Kyoto 606-8502, Japan\\
        E-mail: \email{t-ichi@ruby.scphys.kyoto-u.ac.jp}}

\author{Takashi Z. Nakano\\
	Yukawa Institute for Theoretical Physics
	\& Department of Physics,\\
	Kyoto University,
	Kyoto 606-8502, Japan\\
        E-mail: \email{t-nakano@ruby.scphys.kyoto-u.ac.jp}}

\abstract{
We investigate the QCD phase diagram in the strong coupling limit
by using a newly developed auxiliary field Monte-Carlo (AFMC) method.
Starting from an effective action in the leading order
of the $1/g^2$ and $1/d$ expansion
with one species of unrooted staggered fermion,
we solve the many-body problem exactly
by introducing the auxiliary fields
and integrating out the temporal links and quark fields.
We have a sign problem in AFMC, which is different from the original one
in finite density lattice QCD.
For low momentum auxiliary field modes,
a complex phase cancellation mechanism exists,
and the sign problem is not serious on a small lattice.
Compared with the mean field results,
the transition temperature is found to be reduced by around 10 \%
and the hadron phase is found to be extended in the larger chemical potential
direction by around 20 \%,
as observed in the monomer-dimer-polymer (MDP) simulations.
}

\FullConference{The 30 International Symposium on Lattice Field Theory - Lattice 2012,\\
		June 24-29, 2012\\
		Cairns, Australia}

\begin{document}

\section{Introduction}

QCD phase diagram is expected to have rich structure,
extensively studied in current high-energy heavy ion collision experiments,
and closely related to compact astrophysical objects
and phenomena~\cite{Review}.
In the early universe as well as in heavy ion collisions at collider energies,
the quark gluon plasma evolves to hadronic matter via the crossover transition
at small quark chemical potential $\mu$ and high temperature $T$.
The first order phase transition from hadronic to quark matter
may be realized in cold dense matter such as the neutron star core.
The QCD critical point (CP) connects these crossover
and first order transitions, and characteristic
large fluctuations of the order parameter may be observed
in heavy-ion collisions or in black hole formation processes~\cite{BHCP}.

The phase transition at finite density is, unfortunately,
much less known compared with the low density crossover transition.
The sign problem of lattice QCD at finite $\mu$ makes it difficult
to perform precise calculations of dense cold matter.
The cancellation of the statistical weight given by the fermion determinant
may not be just a technical problem,
as recent studies suggest
that we cannot reach CP in phase quenched simulations~\cite{NoGo}.
Therefore we need to find other methods than the phase quenched simulation
in order to directly sample appropriate configurations
in dense cold matter 
for the discussion of CP and the first order transition.

One of the hopes may be found in the strong coupling lattice QCD.
In the strong coupling region,
we can carry out the link variable integral prior to the fermion integral
for each order of the inverse squared coupling $1/g^2$~\cite{KawamotoSmit}.
The QCD phase diagram has been studied under the mean field approximation in
the strong coupling limit (leading order, $\mathcal{O}(1/g^0)$)~\cite{MF-SCL},
the next-to-leading order (NLO, $\mathcal{O}(1/g^2)$)~\cite{NLO}, 
and the next-to-next-to-leading order (NNLO, $\mathcal{O}(1/g^4)$)~\cite{NNLO}.
We can also take account of the fluctuation effects
beyond the mean field approximation.
We obtain the effective action of quarks after the link variable integral, 
and the fermion integral leads to the partition function
in the form of sum over monomer-dimer-polymer (MDP)
configurations~\cite{KarschMutter}.
The fluctuation effects are found to modify the phase diagram shape moderately:
The transition temperature is shifted to lower $T$ direction by 10-20 \%,
and the hadronic phase expands to higher $\mu$ direction by 20-30 \%~\cite{MDP}.
Until now, MDP simulation has been performed only in the strong coupling
limit, $1/g^2=0$.
Since both finite coupling and fluctuation effects are important
to discuss the QCD phase diagram,
we need to develop a theoretical framework which includes
both of these effects.

In this work, we develop an auxiliary field Monte-Carlo (AFMC) method
for the strong coupling lattice QCD.
In AFMC, we carry out the integral over the auxiliary fields,
which are introduced to decompose the fermion composite product.
In the mean field method, 
auxiliary fields are introduced and assumed to be static and constant.
Thus AFMC is a straightforward extension of the mean field method,
and may be also applicable to finite coupling cases.
We discuss here the QCD phase diagram in the strong coupling limit.

\section{Auxiliary field Monte-Carlo method}

We consider an asymmetric lattice ($a_\tau=a/\gamma$)
of size $L^3\times N_\tau$
for color $\mathrm{SU}(N_c=3)$
with one species of unrooted staggered fermion
in the strong coupling limit (SCL).
Throughout this paper, we work in the lattice unit $a=1$,
where $a$ is the spatial lattice spacing.
The lattice QCD action is given as
\begin{align}
S_\mathrm{LQCD}
&=\frac12 \sum_x \left[
	V^{+}_x - V^{-}_x
		\right]
+\frac{1}{2\gamma} \sum_{x,j} \eta_{j,x}\left[
		 \bar{\chi}_x U_{j,x} \chi_{x+\hat{j}}
		-\bar{\chi}_{x+\hat{j}} U^\dagger_{j,x} \chi_x
		\right]
+\frac{m_0}{\gamma} \sum_{x} M_x
\label{Eq:LQCD}
\ ,\\
V^{+}_x&=e^{\mu/\gamma^2} \bar{\chi}_x U_{0,x} \chi_{x+\hat{0}}
\ ,\quad
V^{-}_x=e^{-\mu/\gamma^2} \bar{\chi}_{x+\hat{0}} U^\dagger_{0,x} \chi_x
\ ,\quad
M_x=\bar{\chi}_x \chi_x
\ ,
\end{align}
where $\chi_x$ and $U_{\nu,x}$ represent the quark field and the link variable,
respectively,
$\eta_{j,x}=(-1)^{x_0+\cdots+x_{j-1}}$ is the staggered sign factor,
and chemical potential $\mu$ is introduced
in the form of the temporal component of a vector potential.
In SCL, we can ignore the plaquette action terms,
which are proportional to $1/g^2$.
By integrating out spatial link variables,
we obtain the SCL effective action~\cite{DKS},
\begin{align}
S_\mathrm{eff}
&=\frac12 \sum_x \left[ V^{+}_x - V^{-}_x \right]
-\frac{1}{4N_c\gamma^2} \sum_{x,j} M_x M_{x+\hat{j}}
+\frac{m_0}{\gamma} \sum_{x} M_x
\ .\label{Eq:Seff}
\end{align}
Here we adopt the effective action in the leading order of
the $1/d$ expansion, where $d$ is the spatial dimension, $d=3$.
The nearest neighbor four fermi interaction,
the second term in Eq.~\eqref{Eq:Seff},
is rewritten in the momentum representation as,
\begin{align}
-\frac{1}{4N_c\gamma^2} \sum_{j,x} M_x M_{x+\hat{j}}
= -\frac{L^3}{4N_c\gamma^2}
	\sum_{\bold{k}, \tau}
	f(\bold{k})\, M_{-\bold{k},\tau}\, M_{\bold{k},\tau}
\ ,\quad
f(\bold{k})=\sum_j \cos\, k_j
\ ,
\label{Eq:Veff}
\end{align}
where the Fourier transformation is defined as
$M_{x=(\bold{x},\tau)}=\sum_{\bold{k}} e^{i\bold{k}\cdot\bold{x}} M_{\bold{k},\tau}$.

We shall now introduce auxiliary fields and decompose the interaction terms.
We apply the extended Hubbard-Stratonovich (EHS)
(or Hubbard-Stratonovich-Miura-Nakano-Ohnishi-Kawamoto)
transformation~\cite{NLO},
\begin{align}
e^{\alpha A B} = \int d\psi d\psi^*
	e^{-\alpha\left[ \psi^* \psi - A \psi - \psi^* B \right]}
\ ,
\end{align}
where $d\psi\,d\psi^*=d\mathrm{Re}\psi\,d\mathrm{Im}\psi$.
By introducing two auxiliary fields simultaneously,
we can bosonize any kind of composite product.
For the interaction term Eq.~\eqref{Eq:Veff},
we find that the positive and negative meson hopping matrix eigenvalues
appear in pair,
$f(\vkb)=-f(\vk)$
with $\bar{\bold{k}}=\bold{k}+(\pi,\pi,\pi)$. 
We refer to the modes with momentum $\vk$ satisfying $f(\vk)>0$ and $f(\vk)<0$ 
as positive and negative modes, respectively.
The bosonization of these terms is carried out as
\begin{align}
&\exp\left\{\alpha f(\bold{k})\left[M_\vkmt M_\vkt
-M_\vkmbt M_\vkbt\right]\right\}
\nonumber\\
&=\int d\sigma_{\vkt}\,d\sigma_{\vkt}^*\,d\pi_{\vkt}\,d\pi_{\vkt}^*
\exp\left\{
-\alpha f(\vk)\left[
 |\sigma_{\vkt}|^2+|\pi_{\vkt}|^2
\right.\right.
\nonumber\\
&\hspace*{4cm}
\left.\left.
+\sigma^*_\vkt M_\vkt
+M_\vkmt \sigma_\vkt 
-i(-)^\tau\pi^*_\vkt M_\vkbt
-i(-)^\tau M_\vkmbt \pi_\vkt 
\right]
\right\}
\ ,
\end{align}
where $\vk$ is chosen to be one of the positive modes, $f(\vk)>0$.
By construction, $\sigma_{\vkt}$ and $\pi_\vkt$ satisfy the relation
$\sigma_\vkmt=\sigma_\vkt^*$ and $\pi_\vkmt=\pi_\vkt^*$.
The bosonized interaction terms are given as
\begin{align}
S_\mathrm{eff}^\mathrm{EHS}
&=\frac{1}{2}\sum_x\left[V_x^+ - V_x^-\right]
 +\sum_x m_x M_x
 +\frac{L^3}{4N_c\gamma^2} \sum_{\vkt, f(\vk)>0}
f(\vk)\left[\left|\sigma_\vkt\right|^2+\left|\pi_\vkt\right|^2\right]
\ ,
\\
m_x
&=
 \frac{m_0}{\gamma}
 +\frac{1}{4N_c\gamma^2} \sum_{j}
	\left[
	 (\sigma+i\varepsilon\pi)_{x+\hat{j}}
	+(\sigma+i\varepsilon\pi)_{x-\hat{j}}
	\right]
\ ,\\
\sigma_x
&= \sum_{\vk,f(\vk)>0} e^{i\bold{k}\cdot\bold{x}}\sigma_\vkt
\ ,\quad
\pi_x
= \sum_{\vk,f(\vk)>0} e^{i\bold{k}\cdot\bold{x}}\pi_\vkt
\ ,
\end{align}
where $\varepsilon_x=(-)^{x_0+x_1+x_2+x_3}$ corresponds to 
$\Gamma_{55}=\gamma_5\otimes\gamma_5$ in the spinor-taste space.
The lattice QCD action Eq.~\eqref{Eq:LQCD} is invariant
under the chiral U(1) transformation,
$\chi_x \to e^{i\varepsilon_x\theta}\chi_x$,
and the chiral transformation mixes $\sigma_\vkt$ and $\pi_\vkt$.
Thus $\sigma_\vkt$ and $\pi_\vkt$ at small $\vk$ are regarded as
the usual chiral ($\sigma$) and Nambu-Goldstone ($\pi$) fields, respectively.

We can carry out the Grassmann and temporal link ($U_0$) integrals
semi-analytically.
\begin{align}
S_\mathrm{eff}^\mathrm{AF}
&=\sum_{\vkt, f(\vk)>0}
  \frac{L^3f(\vk)}{4N_c\gamma^2}
  \left[\left|\sigma_\vkt\right|^2+\left|\pi_\vkt\right|^2\right]
-\sum_x \log\left[
	X_{N_\tau}(\bold{x})^3-2X_{N_\tau}(\bold{x})
	+2\cosh(3N_\tau\mu/\gamma^2)\right]
\ ,
\label{Eq:SeffAF}
\end{align}
where $X_{N_\tau}(\bold{x})$ is a known function of $m_x$~\cite{Faldt},
and we can obtain it by using a recursion formula with $N_\tau$ steps.
When $m_{x=(\bold{x},\tau)}$ is independent of $\tau$,
we obtain $X_{N_\tau}=2\cosh(N_\tau\ \mathrm{arcsinh}\ m_x)$.

We perform Monte-Carlo integral calculations
over the auxiliary fields $(\sigma_\vkt, \pi_\vkt)$
based on the auxiliary field effective action Eq.~\eqref{Eq:SeffAF}.
We refer to this treatment as an auxiliary field Monte-Carlo (AFMC) method.
We have made two approximations
to obtain the effective action Eq.~\eqref{Eq:Seff}
(leading orders of strong coupling expansion and $1/d$ expansion),
whereas no approximations have been invoked
to calculate observables based on $S_\mathrm{eff}$ in Eq.~\eqref{Eq:Seff}.
One of the merit of AFMC in the strong coupling limit
is that the fermion matrix is decomposed into that at each spatial site.
The numerical cost is proportional to the space-time lattice volume
multiplied by the one-dimensional size,
$L^3\times N_\tau \times (L~\mathrm{or}~N_\tau)$.
As in fermion many-body problems in other fields of physics,
unfortunately, we have a sign problem:
The fermion self-energy $m_x$ is complex, and the second term
in the auxiliary field effective action $S_\mathrm{eff}^\mathrm{AF}$
Eq.~\eqref{Eq:SeffAF} contains the imaginary part.
As a result,
the statistical weight $\exp(-S_\mathrm{eff}^\mathrm{AF})$ has a phase,
coming from the negative modes $\pi_\vkt$.
This sign problem is weakened in part by the phase cancellation mechanism.
Since negative modes involve $i\varepsilon_x$,
the phase on one site from low momentum $\pi_\vkt$ modes
is tend to be cancelled by the phase on the nearest neighbor site.
Phase from high momentum modes does not cancel,
but we expect that high momentum modes are less relevant
to long wave phenomena such as the phase transition.
Nevertheless, we demonstrate that AFMC works in a small lattice
such as $4^3\times N_\tau$ and $6^3\times N_\tau$.

\section{Phase diagram in AFMC}

We discuss here the AFMC results of the phase diagram
in the chiral limit $(m_0=0)$ on a small lattice
$4^3\times N_\tau$ and $6^3\times N_\tau$.
Following the arguments in \cite{Bilic}, we assume the temperature
is given as $T=\gamma^2/N_\tau$.

\begin{figure}[bh]
\begin{minipage}[c]{7.5cm}
\PSfig{7.5cm}{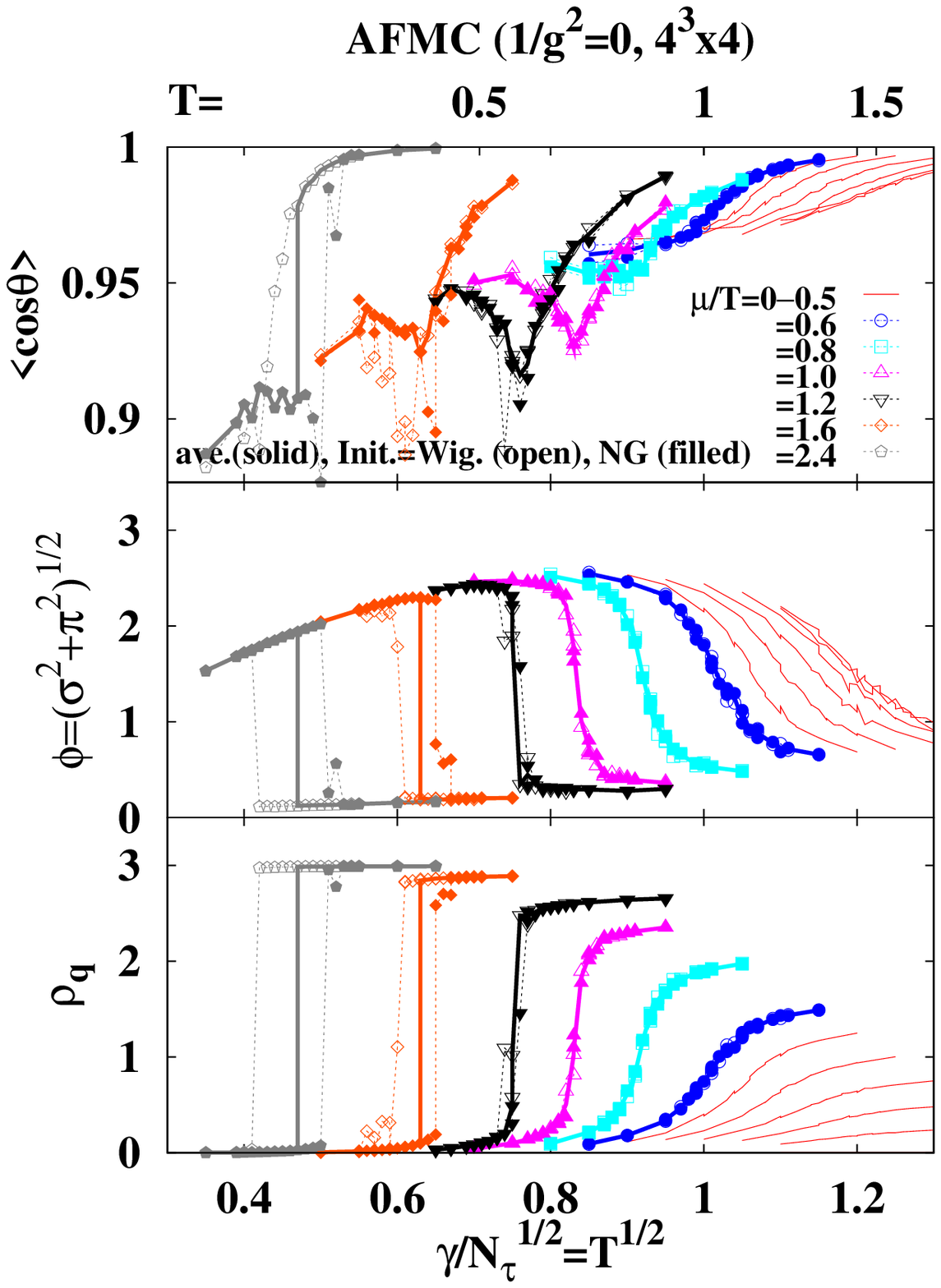}\end{minipage}~\hspace*{1cm}~\begin{minipage}[t]{7.5cm}
\PSfig{7cm}{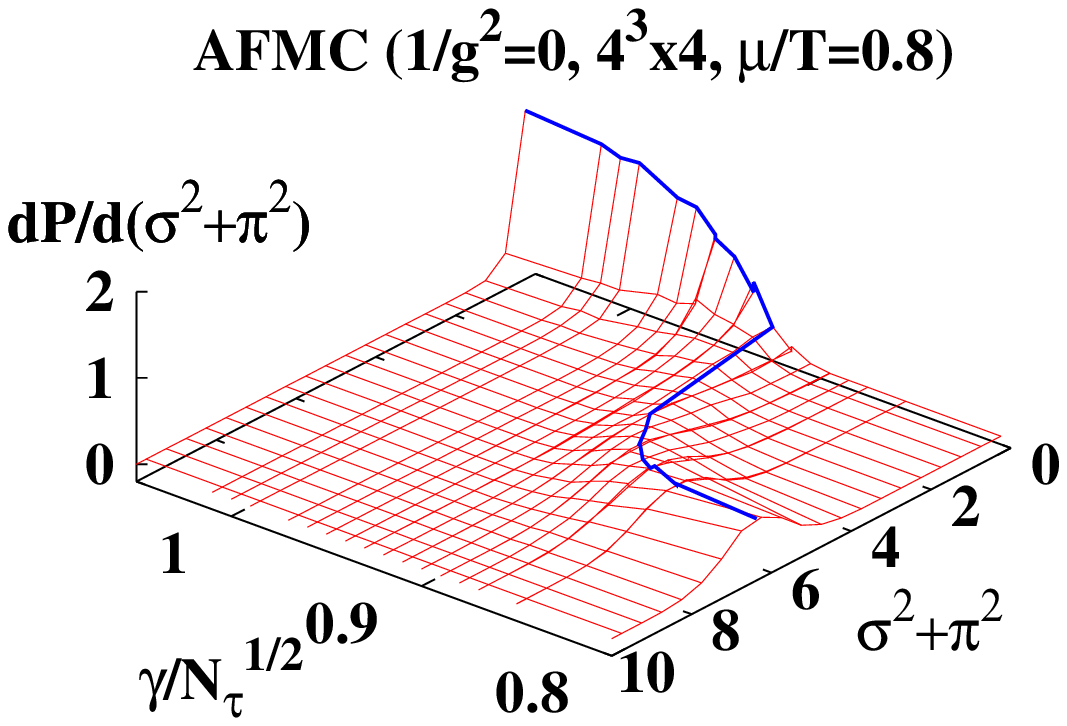}\\
\PSfig{7cm}{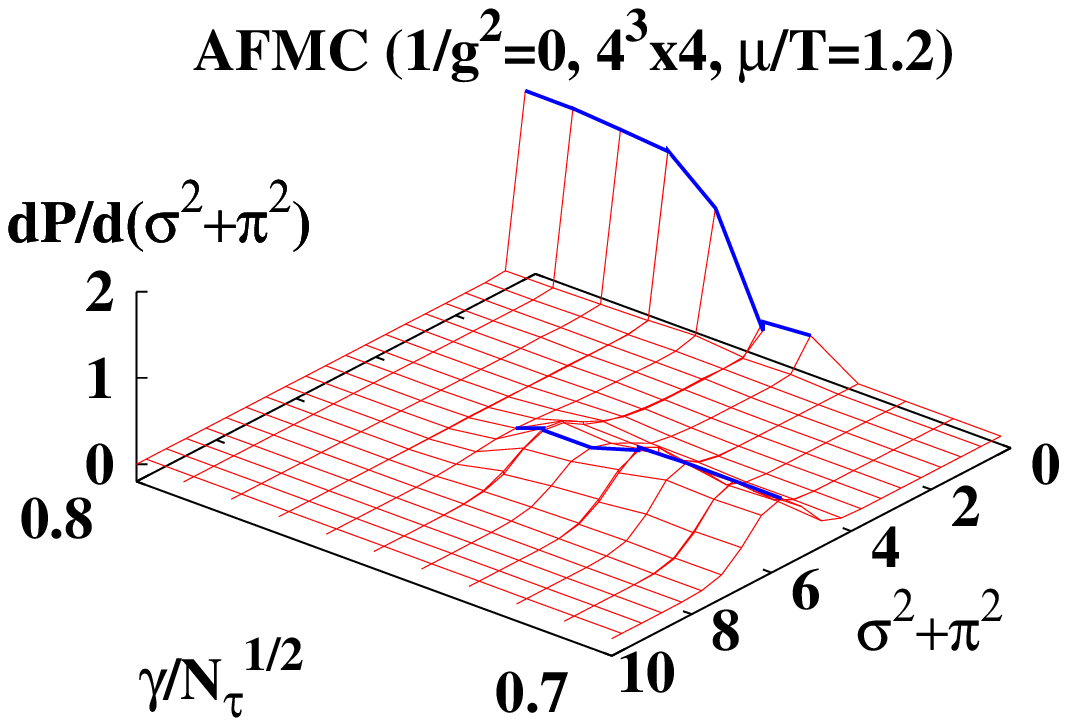}
\end{minipage}
\caption{Average sign factor (top), chiral condensate (middle),
and quark number density (bottom) as functions of temperature
on a $4^4$ lattice.}
\label{Fig:4x4}
\end{figure}

In the left panel of Fig.~\ref{Fig:4x4},
we show the average sign factor $\langle\cos\theta\rangle$,
the root mean square chiral condensate
$\phi=\sqrt{\langle\sigma_{\bold{k}=0,\tau}^2+\pi_{\bold{k}=0,\tau}^2\rangle}$,
and the quark number density $\rho_q$,
as functions of the temperature
on a $4^3\times4$ lattice.
Lines connect fixed $\mu/T$ results.
At small $\mu/T$,
$\phi$ and $\rho_q$ smoothly change around the transition temperature $T_c$.
During this change,
the $\phi^2$ distribution is always single peaked
as shown in the right upper panel of Fig.~\ref{Fig:4x4}.
The peak at a finite $\phi$ at lower $T$
moves to the peak at $\phi=0$ at $T_c$.
This behavior suggests that this transition is a would-be second order
transition, which could be the second order for a large enough volume.
At large $\mu/T$, we find differences between the results starting from
the Wigner phase and the Nambu-Goldstone phase initial conditions
(dashed lines),
as far as the sampling steps are not very large.
In the initial conditions, we set the auxiliary fields static and constant.
We set $\sigma_x=0.01$ and $1$ in the Wigner and Nambu-Goldstone phase
initial conditions, respectively,
and $\pi_x=0$ is taken in both types of initial conditions.
The $\phi^2$ distribution becomes double peaked
as shown in the right bottom panel of Fig.~\ref{Fig:4x4}.
The initial condition dependence suggests the two states are
separated by a high barrier in the effective potential,
and the transition is suggested to be would-be first order.

The average sign factor is large enough, $\langle\cos\theta\rangle \gtrsim 0.9$,
and there is practically no sign problem on a small lattice.
Generally, the average sign factor increases with increasing $T$
for a fixed fugacity,
and increases with increasing $\mu$ except for the transition region.
The global $\mu$ dependence is understood from the form of the effective action.
The imaginary part in $S_\mathrm{eff}^\mathrm{AF}$ comes
from the auxiliary fields, and their effects becomes smaller at larger $\mu$
as we can find in Eq.~\eqref{Eq:SeffAF}.
Around the transition region, finite momentum modes will contribute,
where the phase cancellation mechanism does not work completely,
and the average sign factor is suppressed.

\begin{figure}
\centerline{\PSfig{10cm}{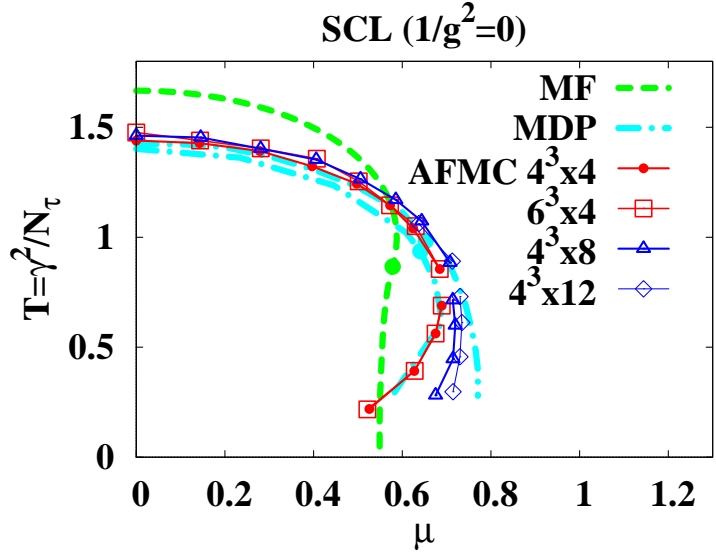}}
\caption{Phase diagram}
\label{Fig:PB}
\end{figure}

We have evaluated the transition temperature
based on the chiral susceptibility peak in the second order region
of $\mu/T$,
and by comparing the effective potential value at the local minima
in the first order region.
In Fig.~\ref{Fig:PB}, we compare the AFMC phase boundary
with that in the mean field approximation~\cite{MF-SCL}
and in the MDP simulation~\cite{MDP}.
As discussed in the previous works using the MDP simulation~\cite{MDP}, 
the transition temperature is found to be reduced at $\mu \simeq 0$
by around 10 \%,
and the hadron phase is found to be extended
in the larger $\mu$ direction at low $T$.

In the strong coupling limit,
the (would-be) first order phase boundary
is insensitive to the spatial lattice size;
the phase boundary on the $6^3\times 4$ lattice
is almost the same as that on the $4^3\times 4$ lattice,
while the transition is found to be shaper on the $6^3\times 4$ lattice.
For larger $N_\tau$ lattices, the transition chemical potential
is found to be larger. Extrapolated results to $N_\tau=\infty$
is consistent with the continuous time MDP simulation~\cite{MDP}.
These observations imply that AFMC can be a promising tool to discuss
finite density lattice QCD, and it can be an alternative method
of the MDP simulation.

\section{Summary}

We have discussed the QCD phase diagram
by using the auxiliary field Monte-Carlo (AFMC) method.
%
Starting from an effective action $S_\mathrm{eff}(\chi,\bar{\chi},U_0)$
of quark and temporal link variables
in the leading order of the $1/g^2$ and $1/d$ expansion
with one species of unrooted staggered fermion,
we obtain an effective action 
$S_\mathrm{eff}^\mathrm{EHS}(\chi,\bar{\chi},U_0,\sigma_\vkt,\pi_\vkt)$
in the bi-linear form of quarks
via the extended Hubbard-Stratonovich transformation.
Integral over the quark and temporal link variables
is carried out analytically,
and the auxiliary field effective action
$S_\mathrm{eff}^\mathrm{AF}(\sigma_\vkt,\pi_\vkt)$
is obtained.
Finally, we integrate out the auxiliary fields
$(\sigma_\vkt,\pi_\vkt)$
using the Monte-Carlo method.
This procedure corresponds to solving the many-body problem
of $S_\mathrm{eff}$ exactly.
%
While we have a sign problem in AFMC,
the problem is not serious on a small lattice partly
due to the phase cancellation mechanism of the low momentum $\pi_\vkt$ modes.
The phase diagram in AFMC is found to be compatible
with that in the monomer-dimer-polymer (MDP) simulations~\cite{MDP}.
Thus the QCD phase diagram in the strong coupling limit in MDP
is confirmed to be correct.

We can extend AFMC to include finite coupling effects
in a straightforward manner.
The sign problem with finite coupling or on a larger lattice,
however, may be more severe.
The present sign problem comes from high momentum negative modes,
then it is an interesting direction to study to integrate out
high momentum auxiliary field modes above a given cutoff
in an approximate manner such as in the saddle point method.
Since the auxiliary fields introduced here are color singlet,
the momentum cutoff does not violate the gauge symmetry.
If the cutoff dependence is small, we hope that it would be possible
to evaluate the phase diagram on a large lattice at finite coupling
with less severe sign problem.

\section*{Acknowledgements}
The authors would like to thank Wolfgang Unger, Philippe de Forcrand, 
Naoki Yamamoto, Kim Splittorff, Jan M. Pawlowski, Mannque Rho, Atsushi Nakamura,
and participants of the YIPQS-HPCI workshop
on "New-type of Fermions on the Lattice"
for useful discussions.
TZN is supported by Grant-in-Aid for JSPS Fellows (No.22-3314 and 10J03314).
This work is supported in part by the Grants-in-Aid for Scientific Research
from JSPS
(Nos.
          (B) 23340067, 
          (B) 24340054, 
          (C) 24540271
),
by the Grants-in-Aid for Scientific Research on Innovative Areas from MEXT
(No. 2404: 24105001, 24105008), 
by the Yukawa International Program for Quark-hadron Sciences,
and by the Grant-in-Aid for the global COE program ``The Next Generation
of Physics, Spun from Universality and Emergence" from MEXT.

\end{document}